\documentclass{article}
\usepackage{epsf}
\makeatletter
\newif\ifpr@pstyle \pr@pstylefalse
\newif\ifnons@qeq  \nons@qeqfalse
\newfont{\fourteencp}{cmcsc10 scaled\magstep2}
\newfont{\titlefont}{cmbx10 scaled\magstep2}
\newfont{\authorfont}{cmcsc10 scaled\magstep1}
\newfont{\fourteenmib}{cmmib10 scaled\magstep2}
	\skewchar\fourteenmib='177
\newfont{\elevenmib}{cmmib10 scaled\magstephalf}
	\skewchar\elevenmib='177
\newfont{\ninemib}{cmmib9} \skewchar\ninemib='177
\makeatletter
\newcommand\nonsequentialeqnum{
        \nons@qeqtrue
	\@addtoreset{equation}{section}
	\def\theequation{\arabic{section}.\arabic{equation}}}
\newif\ifp@bblock  \p@bblocktrue
\newcommand\nopubblock{\p@bblockfalse}
\newcommand\topspace{\hrule height 0pt depth 0pt \vskip}
\newcommand\p@bblock{\begingroup \tabskip=\hsize minus \hsize
	\baselineskip=1.5\ht\strutbox \topspace-2\baselineskip
	\halign to\hsize{\strut ##\hfil\tabskip=0pt\crcr
	\the\Pubnum\crcr\the\date\crcr}\endgroup}
\newcommand\YUKAWAmark{\hbox{
        \ifpr@pstyle\ninemib\else\elevenmib\fi
        Yukawa\hskip1mm Institute\hskip1mm Kyoto \hfill}}
\newtoks\date
\newtoks\Pubnum

\Pubnum={YITP-99-60}
\date={September 1999}
\newcommand{\frontpageskip}{\vspace{12pt plus .5fil minus 2pt}}
\def\@authoraddress{} \def\@title{}
\def\title#1{\gdef\@title{\frontpageskip
	\begin{center}{\titlefont #1}\end{center}\par}}
\def\@author#1{\frontpageskip\par\begin{center}{\authorfont #1}
	\end{center}
	\nobreak}
\def\author#1{\expandafter\def\expandafter\@authoraddress\expandafter
    {\@authoraddress{\@author{#1}}}}
\def\andauthor#1{\expandafter\def\expandafter\@authoraddress\expandafter
    {\@authoraddress{\frontpageskip\centerline{and}\@author{#1}}}}
\def\authors#1{\expandafter\def\expandafter\@authoraddress\expandafter
    {\@authoraddress{\frontpageskip\noindent #1}}}
\def\@address#1{\par\begin{center}{\sl #1}\end{center}\par}
\def\address#1{\expandafter\def\expandafter\@authoraddress\expandafter
    {\@authoraddress{\@address{#1}}}}
\def\andaddress#1{\expandafter\def\expandafter%
    \@authoraddress\expandafter
    {\@authoraddress{\par\centerline{\sl and}\@address{#1}}}}
\renewcommand{\thanks}[1]{\footnote{#1}}
\newlength{\paperbaselineskip}
\def\maketitle{\par
  \begingroup
       \def\thefootnote{\fnsymbol{footnote}}
	\thispagestyle{empty}
        \baselineskip=\paperbaselineskip
	\@maketitle
	\endgroup
	\setcounter{footnote}{0}
	\let\maketitle\relax \let\@maketitle\relax
	\let\@thanks\relax \let\@title\relax
	\let\@title\relax \let\@authoraddress\relax
	\let\thanks\relax}
\def\@maketitle{%
        \ifpr@pstyle\vspace{-1.0cm}\else\vspace{-1.7cm}\fi
	\YUKAWAmark\vskip0.6cm
	\ifp@bblock\p@bblock \else\hrule height 0pt \relax \fi
	\@title
	\@authoraddress
	}
\renewcommand{\abstract}{\par\frontpageskip\centerline{
             \ifpr@pstyle\twelvecp\else\fourteencp\fi Abstract}
	\vspace{8pt plus 3pt minus 3pt}}
\makeatother

\def\Journal#1#2#3#4{{#1} {\bf #2}, #3 (#4)}
                                   
\begin{document}

\title{Theory of Nuclear Reactions Leading to Superheavy Elements}

\author{\vspace{-10mm} 
Y. Abe$^1$, K. Okazaki$^2$, Y. Aritomo$^3$, T. Wada$^2$ and M. Ohta$^2$}

\address{\vspace{-10mm} 
$^1$Yukawa Institute for Theoretical Physics, Kyoto Univ., Kyoto
606-0l, Japan\\
$^2$Department of Physics, Konan Univ., Kobe 658, Japan\\
$^3$Flerov Laboratory of Nuclear Reactions, JINR, Dubna 141980, Russia\\
}

\maketitle

\vspace{-8mm}

\begin{abstract}
Dynamical reaction theory is presented for synthesis of superheavy
elements. Characteristic features of formation and surviving are
discussed, which combinedly determine final residue cross
sections of superheavy elements. Preliminary results on Z=114 are also 
given.
\end{abstract}

\section{Introduction}
Superheavy elements around Z = 114(or 126) and N = 184 have been
believed to exist according to theoretical prediction of stability
given by the shell correction energy in addition to average nuclear
binding energy.\cite{ar}  This means that heavy atomic nuclei with fissility
parameter {\it x} $\geq$ 
1 could be stabilized against fission by a
huge barrier which is resulted in by the additional binding of the
shel1l correction energy around the spherical shape.  In other words, if
superheavy compound nuclei(C.N.) are formed in such high excitation
that closed shell structure is mostly destroyed, they have no barrier 
against fission and thus are inferred to decay very quickly.
Therefore, the point is how to reach the ground state of the
superheavy nuclei, or how to make a soft-landing at them. In order to
minimize fission decays of C.N. or maximize their
survival probabilities, so-called cold fusion reactions have been
used, which succeeded in synthesizing SHEs up to Z = 112.\cite{ho}  They have
the merit of large survival probabilities, but suffer from the demerit 
of small formation probabilities because of the sub-barrier fusion.
On the other hand, so-called hot(warm) fusion reactions have the merit 
of expected large formation probabilities and the demerit of small
survival probabilities due to relatively high excitation of
C.N. formed. 
An optimum condition for large residue cross
sections of SHEs is a balance or a compromise between formation and
survival probabilities as a function of incident energy or excitation 
energy of C.N. formed.\cite{ab}
Therefore, the whole reaction process has to be described, from the
encounter of incident ions to the formation of compound nuclei, and
then to fission decays with residues left of a small probability.
An optimum path has to be searched for over all possible incident
channels including secondary beams becoming available sooner or later.
 Whether the reaction is ``cold'' or ``hot'' fusion is automatically
determined by choosing combinations of projectiles and targets.  The
theoretical framework, thus, has to accommodate both of them.  As is
described in section~\ref{sec:tf}, the strong dissipation of 
energies of nuclear
collective motions, which is well recognized by the fusion
hindrance\cite{sw} and the long fission life,\cite{wa} has to be 
taken into account in
formation process as well as in survival process. 

\section{Theoretical framework}\label{sec:tf}
In reactions of massive systems, it is not clear that the compound
nucleus theory can apply, i.e., that the formation and the decay of
compound nuclei are independent.  But for simplicity, we assume that
is valid, at least approximately.  Then, the residue cross section is
given by the following formula as usual,

\begin{equation}
\sigma_{SHE} = {\pi\over k^{2}}\sum_{J}(2J+1)\cdot
P_{for}^{J}(Ec.m.) \cdot P_{surv}^{J}(E^{*})\label{eq:one} 
\end{equation}

,where P$_{for}^{J}$ and P$_{surv}^{J}$ denote the formation
probability and the survival probability, respectively.

\subsection{Formation probability}
P$_{for}^{J}$ is not equal to a simple transmission coefficient
T$_{J}$(Ec.m.) of an optical potential, nor to a barrier penetration
factor P$_{J}$(Ec.m.) of the combined potential of Coulomb repulsion and 
nuclear attraction. As is well known,\cite{sw} there is the fusion hindrance in
massive systems, roughly those with Z$_{1}$¡¦Z$_{2}$ $\geq$ 1,800,
which could be interpreted by the overcome of so-called conditional
saddle under the strong dissipation.  The necessary additional energy
is so-called extra-push energy.  Therefore, P$_{for}^{J}$ should take
into account both the usual barrier penetration and the dissipative
dynamics over the saddle.  The former is approximated by the
penetration factor for the parabolic barrier, while the latter is
treated by Langevin equation, concerning the collective shape degrees
of freedom of the compound systems starting with the contact
configuration of the projectile and the target.  In one-dimension,
Langevin eq. is given as  

\begin{equation}
m{d^2q\over dt^2} + {\partial V \over \partial q} - \gamma\cdot{dq \over
dt} + R(t) = 0\label{eq:two}
\end{equation}

The friction coefficient $\gamma$ is taken to be equal to
that of so-called one-body wall-and-window formula.\cite{bl}  The last
time-dependent term R(t) in Eq.(\ref{eq:one}) is a random force associated with
the friction force.  The last two terms are a phenomenological
description of effects of the nucleonic degrees of freedom in
excitation, considered as a heat both.  Thus, we assume the
fluctuation-dissipation theorem,

\begin{equation}
<R(t)\cdot R(t')> = \gamma\cdot T\cdot\delta(t-t')
\end{equation}

,where $<           >$ denotes an average over all the possible
realizations of assumed gaussian noise of R and T does the temperature of
the heat bath, i.e., the compound nucleus.  The delta function in the
r.h.s. is due to the Markovian assumption. 
Equivalently, one can use Kramers eq.

\begin{equation}
{\partial f(q, p, t)\over \partial t} =  \lbrace - {\partial \over
\partial q}{p \over m}+{\partial \over \partial p}{\partial V \over
\partial q} + {\partial \over \partial p}(\beta\cdot p + {m\beta T}{\partial 
\over \partial p}) \rbrace f(q,p,t)\label{eq:four} 
\end{equation}

, where p denotes the conjugate momentum to q, and $\beta$ is the
reduced friction coefficient $\gamma$/m.  Eq.(\ref{eq:four}) describes 
a time-evolution1
of distribution function in the phase space, while Eq.(\ref{eq:two}) does
trajectories of Brownian particle.  Actually,  Eq.(\ref{eq:four}) 
was proposed by
Kramers\cite{kr} in order to interpret the fission width proposed by Bohr and
Wheeler, from the dynamical view-point. 

\newpage
\begin{figure}[h]
\begin{center}
   \epsfxsize=16.5pc
   \epsfbox{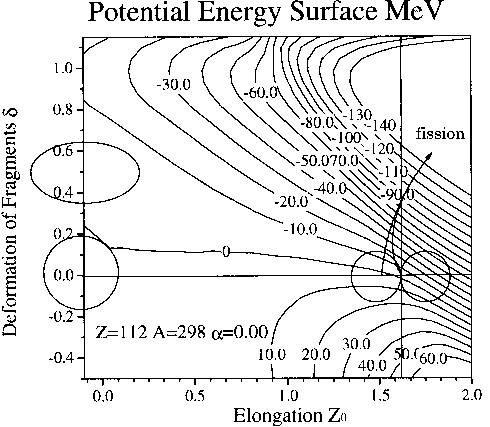}
\caption{ \label{fig:one}}
\end{center}
\end{figure}

\begin{figure}[h]
\begin{center}
    \epsfxsize=16.5pc
    \epsfbox{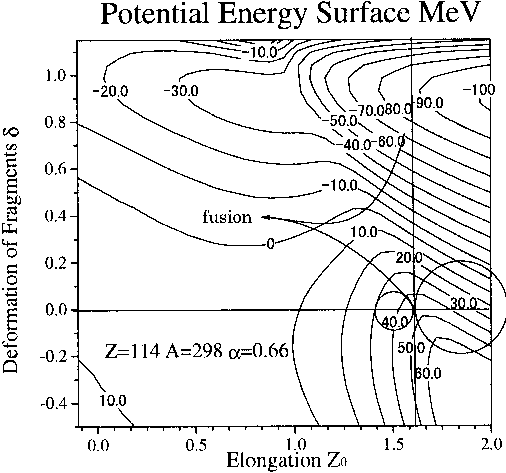}
\caption{ \label{fig:two}}
\end{center}
\end{figure}

In realistic situations, 
processes are in many dimensions including
mass asymmetry degree of freedom etc. in addition to the elongation or 
the separation between fragments. An important case that we will 
discuss below is
that the incident channel is with Z$_1$$\cdot$Z$_2$ $\geq$ 1,800 and 
the compound nucleus is with Z = 114. 
In Fig.~\ref{fig:one}, an example of contour maps of energy surfaces
is given
for the symmetric mass partition (Z$_{1}$¡¦Z$_{1}$$\geq$ 3,000). 
The axes are the elongation (Zo), i.e., the relative
distance between fragments and the deformation where fragment
deformations are assumed to be proportional to their masses.  It is
seen that the contact point of two spheres is located below the
conditional saddle point.  The system has to climb up, governed by the
fluctuation-dissipation dynamics  (the multi-dimentional version of
Eq.(\ref{eq:two}))in order to reach the spherical
configuration of the total system beyond the saddle point.
Apparently, an additional incident energy is required, which
corresponds to the extra-push energy.  On the other hand, if we think
about a very mass-asymmetric incident channel, such as $^{48}$Ca +
$^{244}$Pu
(Z$_{1}$¡¦Z$_{2}$$\simeq$ 1,800),
we may not suffer from the extra energy, because the
contact point of two incident nuclei can be located inside the
conditional saddle point.  Actually, the mass asymmetry parameter
$\alpha$ = 0.66 case is shown in Fig.~\ref{fig:two}, where the contact 
point is slightly
inside and then the composite system evolves mainly over the flat
region around the spherical configuration after touching each other.
This aspect is surely favourable for formation probability. Anyhow,
Langevin trajectories are calculated on the energy surfaces in order
to obtain P$^{J}_{for}$.

\vspace{-2mm}
\subsection{Survival probability}
\begin{figure}[h]
\begin{center}
   \epsfxsize=10pc
   \epsfbox{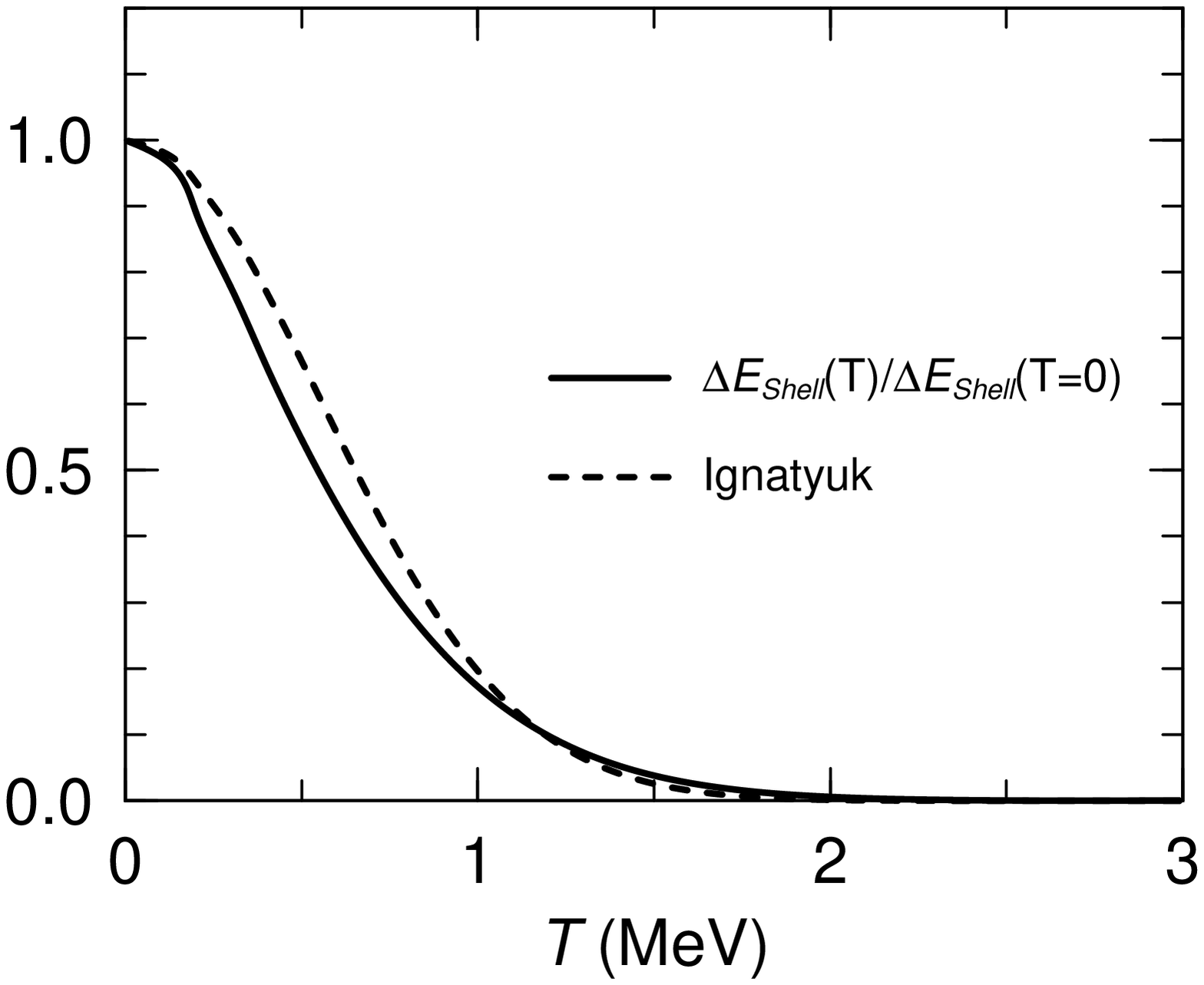}
\caption{\label{fig:three}}
\end{center}
\end{figure}

The survival probability P$_{surv}$ is usually given as follows,

\begin{equation}
P_{surv}(E^{*})={\Gamma_{n}(E^{*})\over \Gamma_{n}(E^{*}) 
	+ \Gamma_{f}(E^{*})}\label{eq:five}
\end{equation}

, assuming that fission and neutron emission are dominant decay 
modes.   The neutron emission width $\Gamma_{n}$ can be calculated by
Weiskopf formula, while the fission width by Bohr-Wheeler one or
Kramers' stationary limit.\cite{wa}$^,$\cite{kr}  If E$^{*}$ is high 
enough for more than one
neutron emission, then the expression at r.h.s. of Eq.(\ref{eq:five}) is
repeatedly  used for sequential E$^{*}$ and factorized to obtain the
final P$_{surv}$.  Anyhow, this method is valid only for cases with the
fission barrier height being larger than the temperature T.  In SHE
compound nuclei, especially in hot fusion reactions, there might be no 
fission barrier, because the barrier is given by the shell correction
energy around the spherical configuration  or somewhere in deformed
region which is expected to melt in high excitation. The potential
landscape changes depending on the temperature T, i.e., on neutron
emission.  Then, a dynamical treatment is called for.  In other words, 
final residue cross sections of SHE are to be determined by a
competition between  time scale of fission and that of restoration
of the shell correction energy due to cooling by neutron evaporation. 
Therefore,
a speed of cooling and a temperature dependence of the shell
correction energy are crucial.  
The latter is shown in Fig.~\ref{fig:three}, the
energy being normalized by the maximum value, i.e., that of 
T = 0.  The calculations are made by the use

\noindent
of single particle levels in a
Saxon-Wood finite potential well.   The present results are in good
agreement with the prescription proposed by Ignatyuk.\cite{ig} 
Using  this
temperature dependence and cooling obtained by the statistical code for 
evaporation, we obtain potential energy curves as a function of
time. With this time-dependent potential, we calculate fission process
with Eq.(\ref{eq:two}) or (\ref{eq:four}) or its simplified version
and obtain a small fraction 
of probability captured by the pocket made by the restored shell
correction energy, which gives a final survival probability P$_{sur}$.

\begin{figure}[h]
\begin{center}
   \epsfxsize=10pc
   \epsfbox{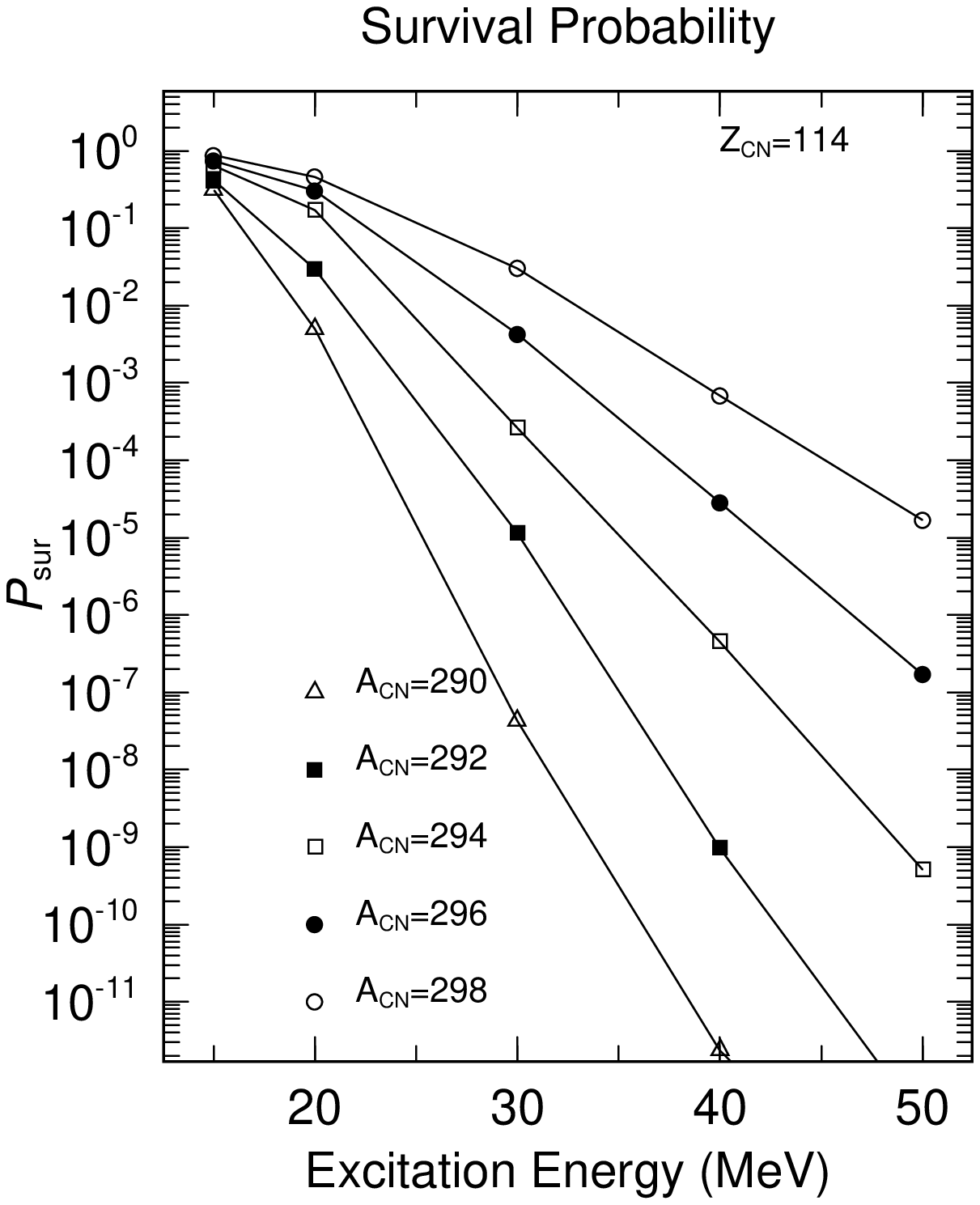}
\caption{\label{fig:four}}
\end{center}
\end{figure}

In Fig.~\ref{fig:four}, P$_{sur}$'s with J=10 are shown for several 
isotopes of compound nuclei with
Z=114, as a function of initial excitation energy.  We readily see
that  P$_{sur}$ depends strongly on excitation energy as well as on 
mass number, i.e., neutron number
of initial compound nucleus.  This is due to differences in cooling
speed due to different neutron separation energies B$_n$'s.  Thus, we
have to form a compound nucleus which has as small B$_n$ as possible,
i.e., as large neutron number N as possible in order to obtain large
survival probabilities. In cases with small N such as GSI experiments, 
higher excitation is extremely unfavored, thus ``cold'' fusion process 
is used.
Note that the results in Fig.~\ref{fig:four} are at the time 2,000 $\times$ 
10$^{-21}$ sec and therefore they should be reduced by a reduction factor
due to the last neutron emission for the final survival probabilitites.

\section{Preliminary results and discussion}
\begin{figure}[h]
\begin{center}
   \epsfxsize=9pc
   \epsfbox{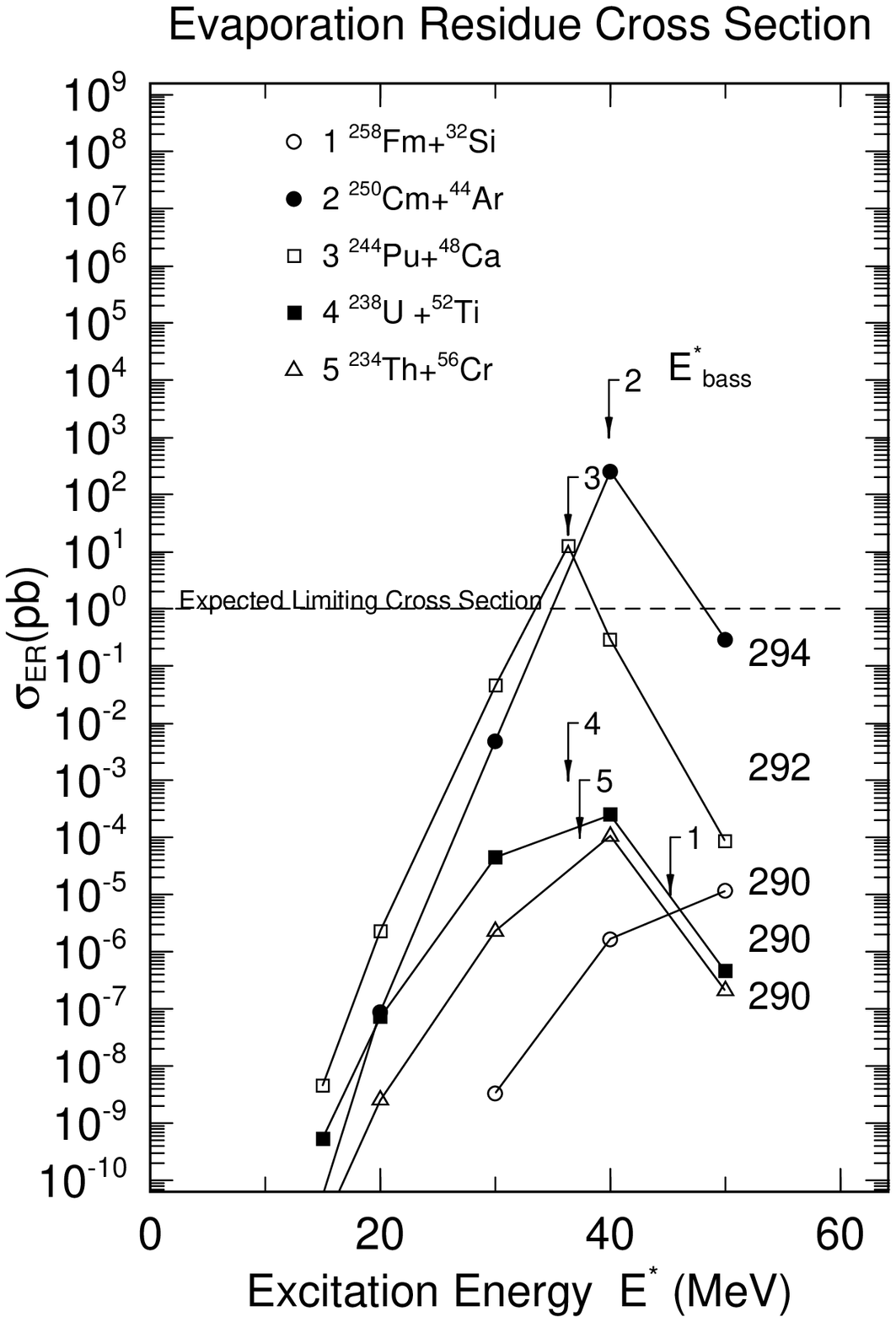}
\caption{\label{fig:five}}
\end{center}
\end{figure}

In brief, using WKB
approximation for tunneling and 3-dimensional Langevin equation for 
subsequent dynamical evolutions of the total system,  we calculated
formation probabilities, while one-dimensional Smoluchowski equation
which is the approximation of Eq.(\ref{eq:four})
was used for fission decay or survival probabilities.  Evaporations
were taken into account in statistical model.  Results are shown in
Fig.~\ref{fig:five} over possible incident channels leading to Z=114. 
Existences of 
the optimum enegies in various channels are clearly seen, which
resulted in by the compromise of the two factors. Rapid increases in
the left hand side are due to barrier penetration and dissipative dynamical
evolutions of shape degrees of freedom, while decreases in the right
hand side due to energy dependences of survival probabilities against
fission. It is worth
to notice that the recent Dubna experiment observed an event which
would be a signature for Z=114 element, at the energy predicted here
to be most favourable in the incident channel $^{48}$Ca + $^{244}$Pu.[9]

\begin{figure}[h]
\begin{center}
   \epsfxsize=15.2pc
   \epsfbox{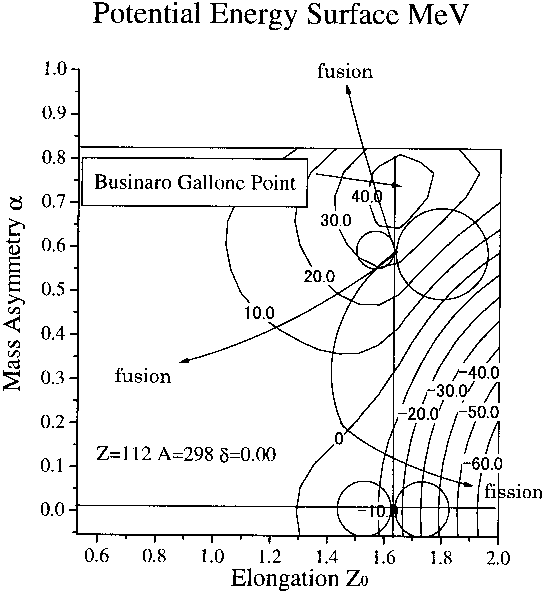}
\caption{ \label{fig:six}}
\end{center}
\end{figure}

As for a relation to so-called dinucleus system concept,\cite{ad} it would be
helpful to have a look on contour map of energy surface in axes of the
elongation (Zo), 
and their mass-asymmetry ($\alpha$), which is shown in
Fig.~\ref{fig:six}.  
There is a Businaro-Gallone peak at the upper right corner.
Touching points with asymmetries $\alpha$ = 0.0 (symmetric case) and
$\alpha$ = 0.66 (close to $^{48}$Ca + $^{244}$Pu case) are also shown,
schematically, It should be noted here that the top horizontal line
and the left vertical line both describe approximately the spherical
configuration of the total system, thereby should converge to one
point in a more realistic representation. The model based on the 
dinucleus system concept gives
fusion probabilities over the Businaro-Gallone peak, so it evaluates evolutions
along the vertical direction in Fig.~\ref{fig:six}, while the 3-D
Langevin calculations take into account
evolutions over all possible paths dynamically including the
horizontal direction as well.

\vspace{-12mm}
\begin{thebibliography}{99}
\vspace{-2mm}
\baselineskip3mm
\bibitem{ar} see for example, 
	P. Armbruster, \Journal{Ann. Rev. Nucl. Sci.}{35}{135}{1985}.
\bibitem{ho} S. Hofmann et al., \Journal{Z. Phys.}{A354}{229}{1996}.
\bibitem{ab} Y. Abe et al., \Journal{J. Phys.}{G23}{1275}{1997}.\\
	Y. Aritomo et al., \Journal{Phys.Rev.}{C55}{R1011}{1997}, and
	{\it ibid}\Journal{}{C59}{796}{1999}.\\
	T. Wada et al., Proc. DANF98, Slovakia, Oct. 1998.
\bibitem{sw} W.J. Swiatecki, \Journal{Nucl. Phys.} {A376}{275}{1982}.\\
	S. Bjornholm and W.J. Swiatecki,
	 \Journal{Nucl. Phys.}{A391}{471}{1982}.
\bibitem{wa} T. Wada, Y. Abe and N. Carjan, \Journal{Phys. Rev. Lett.}{70}
	{3538}{1993}\\
	Y. Abe et al., \Journal{Phys. Reports}{C275}{49}{1996}.
\bibitem{bl} J. Blocki et al., \Journal{Ann. Phys. (N.Y.)} {105}{427} {1997}.
\bibitem{kr} H.A. Kramers, \Journal{Physica} {VII4}{284}{1940}.
\bibitem{ig} A.V. Ignatyuk et al.,
	 \Journal{Sov. J. Nucl. Phys.}{21}{255}{1975}.
\bibitem{og} Yu. Oganessian et al., preprint JINR, E7-99-53.
\bibitem{ad} G.G. Adamian et al., \Journal{Nucl. Phys.}{A627}{361}{1997}.
\end{thebibliography}
\end{document}